\documentclass{article}
\usepackage{graphicx}
\usepackage{pdflscape}

\usepackage{amsmath}
\usepackage{a4wide}
\usepackage[a4paper]{geometry}
\usepackage{etoolbox}
\usepackage{lscape}
\usepackage{rotating}
\usepackage{graphicx}% Include figure files
\usepackage{dcolumn}% Align table columns on decimal point
\usepackage{bm}% bold math
\usepackage{hyperref}
\usepackage{xcolor}
\usepackage[utf8]{inputenc}
\usepackage[T1]{fontenc}
\usepackage{mathptmx}
\usepackage{amsmath}

\newcommand{\ket}[1]{\left|#1\right\rangle}
\newcommand{\bra}[1]{\left\langle#1\right|}

\begin{document}

\title{Analytical view on N bodies interacting with quantum cavity in tight-binding model}
\author{Krzysztof Pomorski$^{1,2}$} % \newline
%%\author{1:School of Computer Science, University College Dublin \newline
%%2:Quantum Hardware Systems: $www.quantumhardwaresystems.com$ }
 %, Dublin 4, Ireland }
%\author{Krzysztof Pomorski} % Write as First name Surname
% \email[Corresponding author: ]{Email: kdvpomorski@gmail.com}
%% \affiliation{
%% School of Computer Science, University College Dublin %, Dublin 4, Ireland.% Force line breaks with \\ if necessary
%%}
%\affiliation{
%  School of Electrical and Electronic Engineering, University College Dublin,}
%\affiliation{
%  Quantum Hardware Systems: $www.quantumhardwaresystems.com$,}

  %Dublin 4, Ireland.% Force line breaks with \\ if necessary
%%}
%}%
%\affiliation{You would list an author's second affiliation (if applicable) here.}

\date{ $^1$School of Computer Science, University College Dublin \newline
       $^2$Quantum Hardware Systems\footnote{Webpage:\url{www.quantumhardwaresystems.com}}
 \newline \newline
\today } % It is always \today, today, but any date may be explicitly specified
              % Not printed for conference proceedings
\maketitle

\begin{abstract}
Dynamics of N bodies interacting with quantum cavity is presented. The rotating frame approximation is not used and obtained solutions are the most basic
in the framework of generalized Jaynes-Cummings tight-binding model. All presented solutions are entirely analytical and are expressed in terms of elementary functions. Presented scheme can easily be generalized into N bodies (qubits) with M energetic levels interacting with quantum electromagnetic cavity with K energetic levels. Presented framework can be used for construction of software modeling quantum communication between semiconductor single-electron position-based qubits.
\end{abstract}

\section{Technological motivation
%\label{sec:level1}First-level heading:\protect\\ The line
%break was forced \lowercase{via} \textbackslash\textbackslash
}
\begin{figure}
    \centering
    \includegraphics[width=0.9\columnwidth]{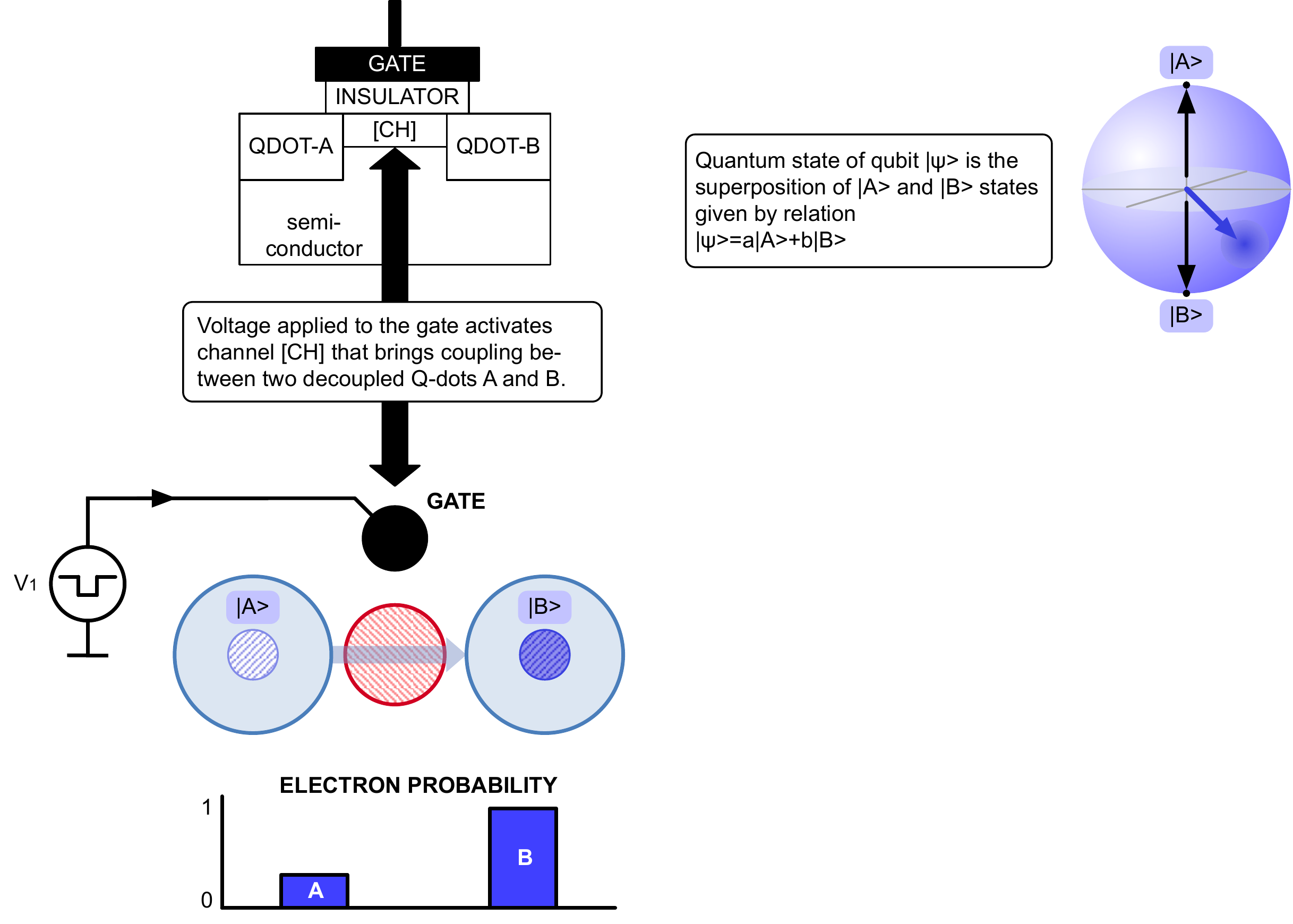} %{QUBIT_SUMMARY11QM.png}
    \caption{Basic concept of position based qubit \cite{Pomorski_spie} and its correspondence to Bloch sphere \cite{Nbody}.}
 \label{central1}
%\label{central}
\includegraphics[width=4.0in]{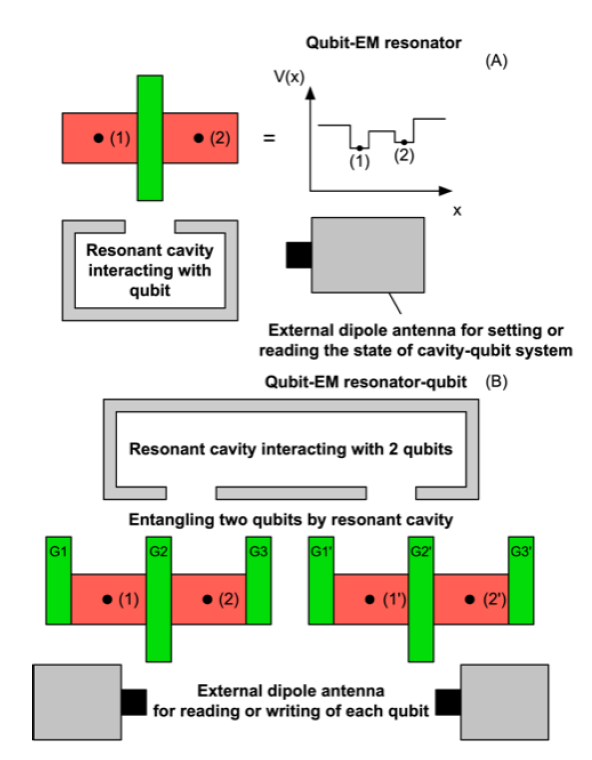} %%%%figs/position_based_Qubit_in_RF_signaLLQ.png} %%{position_based_Qubit_in_RF_signalQ.png} %{position_based_Qubit_in_RF_signal.png}
\caption{Position based qubit in RF field (A) and position based qubits placed at high distance interlinked by waveguide (B) \cite{Nbody}. Physical states of qubits are controlled by voltages applied to the gates G1, G2, G3 and G1', G2', G3'.}
\label{central2}
\end{figure}
Single-electron semiconductor devices are now actively researched for their potential in realizing quantum computers (QC), and especially for implementing single-chip CMOS QCs that are fully integrated with their surrounding electronics \cite{Bashir19}. They were studied by Fujisawa \cite{Fujisawa}, Petta \cite{Petta}, Leipold \cite{Dirk}, Giounanlis \cite{Panos}, Pomorski \cite{Pomorski_spie},\cite{SEL}, \cite{Cryogenics},\cite{qinternet}, \cite{Nbody} and many others. On the other hand, one of the most successful models in condensed matter physics is Hubbard model and its special case known as tight-binding model \cite{Spalek}. We consider a two-energy-level system of position-based (a.k.a. charge) qubit in a tight-binding approach that is a predecessor of Hubbard model as depicted in Fig.\,\ref{central1}. %\,\ref{central}.
%\begin{equation}
%    \label{simple_equation}
%    \alpha = \sqrt{ \beta }
%\end{equation}

The Hamiltonian of this system
is given as $\hat{H}(t)_{[x=(x_1,x_2)]}=$
\begin{eqnarray}
\label{simplematrix}
%=\hat{H}(t)=
%\hat{H}(t)_{[x=(x_1,x_2)]}=
\begin{pmatrix}
E_{p1}(t) & t_{s12}(t)=|t_{s12}|e^{+i\alpha(t)} \\
t_{s12}^{\dag}(t)=|t_{s12}|e^{-i\alpha(t)} & E_{p2}(t)
\end{pmatrix}
=(E_1(t)\ket{E_1}_t \bra{E_1}_t+E_2(t)\ket{E_2}\bra{E_2})_{[E=(E_1,E_2)]}.
\end{eqnarray}
The $\hat{H}(t)$ Hamiltonian's eigenenergies $E_1(t)$ and $E_2(t)$, with $E_2(t)>E_1(t)$, are given,
with $t_{s12}(t)=t_{sr}(t)+it_{si}(t)$, as:
\begin{eqnarray}
E_1(t)= \left(-\sqrt{\frac{(E_{p1}(t)-E_{p2}(t))^2}{4}+|t_{s12}(t)|^2}+\frac{E_{p1}(t)+E_{p2}(t)}{2}\right), \nonumber \\
E_2(t)= \left(+\sqrt{\frac{(E_{p1}(t)-E_{p2}(t))^2}{4}+|t_{s12}(t)|^2}+\frac{E_{p1}(t)+E_{p2}(t)}{2}\right),
\end{eqnarray}
and energy eigenstates $\ket{E_1(t)}$ and $\ket{E_2(t)}$ have the following form
\begin{eqnarray}
\ket{E_1,t}=
\begin{pmatrix}
\frac{(E_{p2}(t)-E_{p1}(t))+\sqrt{\frac{(E_{p2}(t)-E_{p1}(t))^2}{2}+|t_{s12}(t)|^2}}{-i t_{sr}(t)+t_{si}(t)} \\
-1
\end{pmatrix}=
\frac{(E_{p2}(t)-E_{p1}(t))+\sqrt{\frac{(E_{p2}(t)-E_{p1}(t))^2}{2}+|t_{s12}(t)|^2}}{-i t_{sr}(t)+t_{si}(t)}\ket{x_1}-\ket{x_2},\nonumber \\
\ket{E_2,t}=
\begin{pmatrix}
\frac{-(E_{p2}(t)-E_{p1}(t))+\sqrt{\frac{(E_{p2}(t)-E_{p1}(t))^2}{2}+|t_{s12}(t)|^2}}{t_{sr}(t) - i t_{si}(t)} \\
1
\end{pmatrix}=
\frac{-(E_{p2}(t)-E_{p1}(t))+\sqrt{\frac{(E_{p2}(t)-E_{p1}(t))^2}{2}+|t_{s12}(t)|^2}}{t_{sr}(t) - i t_{si}(t)}\ket{x_1}+\ket{x_2}. \nonumber \\
\end{eqnarray}
The last expressions can be written in a compact form
\begin{eqnarray}
\begin{pmatrix}
\ket{E_1,t} \\
\ket{E_2,t} \\
\end{pmatrix}=
\hat{S}_{2 \times 2}
\begin{pmatrix}
\ket{x_1} \\
\ket{x_2} \\
\end{pmatrix}
=
\begin{pmatrix}
\frac{(E_{p2}(t)-E_{p1}(t))+\sqrt{\frac{(E_{p2}(t)-E_{p1}(t))^2}{2}+|t_{s12}(t)|^2}}{-i t_{sr}(t)+t_{si}(t)} & -1 \\
\frac{-(E_{p2}(t)-E_{p1}(t))+\sqrt{\frac{(E_{p2}(t)-E_{p1}(t))^2}{2}+|t_{s12}(t)|^2}}{t_{sr}(t) - i t_{si}(t)} & 1 \\
\end{pmatrix}
\begin{pmatrix}
\ket{x_1} \\
\ket{x_2} \\
\end{pmatrix}.
\end{eqnarray}
Setting $t_{si}(t)=1, t_{sr}(t)=0$ and $E_{p1}(t)=E_{p2}(t)=E_p$ we obtain
%\begin{eqnarray}
$\begin{pmatrix}
\ket{E_2}_n \\
\ket{E_1}_n \\
\end{pmatrix}=\frac{1}{\sqrt{2}}
\begin{pmatrix}
1 &  +1 \\
1 &  -1 \\
\end{pmatrix}
\begin{pmatrix}
\ket{x_2} \\
\ket{x_1} \\
\end{pmatrix}$
%\end{eqnarray}
which brings Hadamard matrix as relating q-state in the position base and in the energy base, where $\ket{E_{1(2)}}_n=\frac{1}{\sqrt{2}}\ket{E_{1(2)}}$.
If we associate logic state 0 with occupancy of node 1 (spanned by $\ket{x_1}$) and logic state 1 with occupancy of node 2 spanned by $\ket{x_2}$, then Hadamard operation on logic state $0$ brings occupancy of $E_2$ (so it is spanned by $\ket{E_2}$) and Hadamard operation on
logic state $1$ brings the entire occupancy of energy level $E_1$ (that is spanned by $\ket{E_1}$).

It shall be underlined that in the most simple case of position-based qubit $E_{p1}=E_{p2}=E_p=\rm const_1$ and $t_{s12}=|t|=\rm const_2$ and we obtain
%
%%\begin{eqnarray}
$\ket{\psi(t)}=\frac{1}{\sqrt{2}}(c_{E_1}e^{\frac{E_1}{\hbar}t}+c_{E2}e^{\frac{E_2}{\hbar}t})\ket{x_1}
+\frac{1}{\sqrt{2}}(-c_{E1}e^{\frac{E_1}{\hbar}t}+c_{E_2}e^{\frac{E_2}{\hbar}t})\ket{x_2}$.
%%\end{eqnarray}
%
It implies an oscillation of probabilities for the electron presence at node 1 (quantum logical 0) and 2 (quantum logical 1) with frequency $2|t|=E_2-E_1$, where $|c_{E_1}|^2 (|c_{E_2}|^2)$ is the probability for the quantum state to be in the ground (excited) state. % denoted by $E_1 (E_2)$. %%% $$.
It is possible to determine the qubit state under any evolution of two eigenergies $E_1(t)$ and $E_2(t)$ that are dependent on $E_{p1}(t), E_{p2}(t), t_{s12}(t)=t_{sr}(t)+t_{si}(t)i$. Simply, we have the state at any time instant given by \begin{eqnarray}
\ket{\psi_t}=e^{\int_{t_0}^{t}\frac{1}{\hbar i}\hat{H}(t')dt'}\ket{\psi_{t_0}}=\hat{U}(t,t_0)\ket{\psi_{t_0}}=
\begin{pmatrix}
e^{\frac{1}{\hbar i}\int_{t_0}^{t}E_1(t')dt'}, 0 \\
0 & e^{\frac{1}{\hbar i}\int_{t_0}^{t}E_2(t')dt'} \\
\end{pmatrix}\ket{\psi_{t_0}},
\end{eqnarray}
We notice that in case of qubit the evolution operator is given as
%\begin{strip}
\begin{eqnarray}
\hat{U}(t,t_0)= %\nonumber \\
\begin{pmatrix}
e^{\frac{1}{\hbar i}\int_{t_0}^{t}\left(-\sqrt{\frac{(E_{p1}(t')-E_{p2}(t'))^2}{4}+|t_{s12}(t')|^2}+\frac{E_{p1}(t')+E_{p2}(t')}{2}\right)dt'} & 0 \\
0 & e^{\frac{1}{\hbar i}\int_{t_0}^{t}\left(+\sqrt{\frac{(E_{p1}(t')-E_{p2}(t'))^2}{4}+|t_{s12}(t')|^2}+\frac{E_{p1}(t')+E_{p2}(t')}{2}\right)dt'} \\
\end{pmatrix},
\end{eqnarray}
\begin{eqnarray}
%%%\nonumber \\
\ket{\psi_t}=c_{e1}(t_0)e^{\frac{1}{\hbar i}\int_{t_0}^{t}\left(-\sqrt{\frac{(E_{p1}(t')-E_{p2}(t'))^2}{4}+|t_{s12}(t')|^2}+\frac{E_{p1}(t')+E_{p2}(t')}{2}\right)dt'}\ket{E_1(t)}+ \nonumber \\ +c_{e2}(t_0)e^{\frac{1}{\hbar i}\int_{t_0}^{t}\left(+\sqrt{\frac{(E_{p1}(t')-E_{p2}(t'))^2}{4}+|t_{s12}(t')|^2}+\frac{E_{p1}(t')+E_{p2}(t')}{2}\right)dt'}\ket{E_2(t)}=, \nonumber \\
=c_{e1}(t_0)e^{\frac{1}{\hbar i}\int_{t_0}^{t}\left(-\sqrt{\frac{(E_{p1}(t')-E_{p2}(t'))^2}{4}+|t_{s12}(t')|^2}+\frac{E_{p1}(t')+E_{p2}(t')}{2}\right)dt'}
\begin{pmatrix}
\frac{((E_{p2}(t)-E_{p1}(t))+\sqrt{\frac{(E_{p2}(t)-E_{p1}(t))^2}{2}+|t_{s12}(t)|^2})e^{i phase(t_{s12}(t))}i)}{\sqrt{|t_s(t)|^2+((E_{p2}(t)-E_{p1}(t))+\sqrt{\frac{(E_{p2}(t)-E_{p1}(t))^2}{2}+|t_{s12}(t)|^2})^2}} \\
\frac{-|t_s(t)|)}{\sqrt{|t_s(t)|^2+((E_{p2}(t)-E_{p1}(t))+\sqrt{\frac{(E_{p2}(t)-E_{p1}(t))^2}{2}+|t_{s12}(t)|^2})^2}}\\
\end{pmatrix}_{x}+\nonumber \\
+c_{e2}(t_0)e^{\frac{1}{\hbar i}\int_{t_0}^{t}\left(+\sqrt{\frac{(E_{p1}(t')-E_{p2}(t'))^2}{4}+|t_{s12}(t')|^2}+\frac{E_{p1}(t')+E_{p2}(t')}{2}\right)dt'}
\begin{pmatrix}
\frac{(-(E_{p2}(t)-E_{p1}(t))+\sqrt{\frac{(E_{p2}(t)-E_{p1}(t))^2}{2}+|t_{s}(t)|^2})e^{-iphase(t_{s12}(t))}}{\sqrt{|t_s|^2+(-(E_{p2}(t)-E_{p1}(t))+\sqrt{\frac{(E_{p2}(t)-E_{p1}(t))^2}{2}+|t_{s12}(t)|^2})^2}} \\
\frac{+|t_s(t)|}{\sqrt{|t_s|^2+(-(E_{p2}(t)-E_{p1}(t))+\sqrt{\frac{(E_{p2}(t)-E_{p1}(t))^2}{2}+|t_{s12}(t)|^2})^2}} \\
\end{pmatrix}_{x}.\nonumber \\
\end{eqnarray}
%\end{strip}
Here, $c_{e1}(t_0)$ and $c_{e2}(t_0)$ describe the qubit in the energy representation at the initial time $t_0$, so $|c_{e1}(t_0)|^2+|c_{e2}(t_0)|^2=1$. Such presented evolution of position-based qubit is under the circumstances of small adiabatic changes in $t_s(t)$ and in
$E_{p1}(t)$, $E_{p2}(t)$. It is not the case of a qubit subjected to the rapid AC field
that will support the existence of resonant states \cite{Pomorski_spie}.

\section{Position-based qubit interaction with quantum electromagnetic cavity}
Jaynes-Cummings Hamiltonian is referring to the following interaction of electromagnetic field with dipole expressed by formula
\begin{equation}
H_{qubit-QEC}=\vec{E}\cdot \vec{d}.
\end{equation}
It is natural to act in the energy eigenbases for quantum electromagnetic cavity and in position bases for qubit. The difference of electrostatic charge
between right and left quantum dot can be accounted by the operator expressed by matrix in the following way as
\begin{equation}\hat{P}_{charge-difference,R-L}=
\begin{pmatrix}
-1 & 0 \\
0 & 1
\end{pmatrix}
\end{equation}
acting on the Wannier function eigenbases of single-electron device.
Therefore electric dipole of 2 coupled dot quantum system is expressed as
\begin{equation}
\hat{d}=\frac{e(x_2-x_1)}{2}
\begin{pmatrix}
-1 & 0 \\
0 & 1
\end{pmatrix},
\end{equation}
where $x_2-x_1$ accounts for distance between 2 centers of coupled quantum dots.
Total Hamiltonian accounts for energy of quantum electromagnetic cavity (QEC), energy of position based qubit and for interaction between quantum electromagnetic cavity and position based qubit and is of the following form
\begin{equation}
H=H_{qubit}+H_{QEC}+H_{qubit-QEC}.
\end{equation}
It leads to the following Hamiltonian
%\tiny
\small
\begin{eqnarray}
    \label{simple_equation}
    H = I_{qed} \times H_{qubit} + H_{qed} \times I_{qubit} + H_{qed-qubit}= \nonumber \\
    =
\begin{pmatrix}
1 & 0 \\
0 & 1
\end{pmatrix} \times
\begin{pmatrix}
E_{p1} & t_s \\
t_s & E_{p2}
\end{pmatrix}
+ % \nonumber \\
\begin{pmatrix}
E_{cav1} & 0 \\
0 & E_{cav2}
\end{pmatrix} \times
\begin{pmatrix}
1 & 0 \\
0 & 1
\end{pmatrix}+ \nonumber \\
(x_2-x_1)e
\begin{pmatrix}
E_{f1}(t) & 0 \\
0 & E_{f2}(t)
\end{pmatrix} \times
\begin{pmatrix}
-1 & 0 \\
0 & 1
\end{pmatrix}= \nonumber \\
%\end{eqnarray}
%\tiny
%\begin{eqnarray}
=
\begin{pmatrix}
E_{p1} & t_s & 0 & 0 \\
t_s^{*} & E_{p2} & 0 & 0 \\
0 & 0 & E_{p1} & t_s \\
0 & 0 & t_s^{*} & E_{p2}
\end{pmatrix}+
\begin{pmatrix}
E_{cav1} & 0 & 0 & 0 \\
0 & E_{cav1} & 0 & 0 \\
0 & 0 & E_{cav2} & 0 \\
0 & 0 & 0 & E_{cav2}
\end{pmatrix}+ \nonumber \\
%\end{eqnarray}
%\tiny
%\begin{eqnarray}
%%%\nonumber \\
 +\frac{e(x_2-x_1)}{2}
\begin{pmatrix}
-E_{f1} & 0  & 0 & 0 \\
0 & E_{f1} & 0 & 0 \\
0 & 0  & -E_{f2} & 0 \\
0 & 0  & 0 & +E_{f2}
\end{pmatrix}= \nonumber \\
%\end{eqnarray}
%\begin{eqnarray}=
\begin{pmatrix}
E_{c1} -E_{f1q}(t)+E_{p1} & t_s  & 0 & 0 \\
t_s^{*} & E_{c1} +E_{f1q}+E_{p2} & 0 & 0 \\
0 & 0  & E_{c2} -E_{f2q}+E_{p1} & t_s \\
0 & 0  & t_s^{*} & +E_{c2}+E_{f2q}+E_{p2}
\end{pmatrix}= \nonumber \\
%\end{eqnarray}
%\normalsize
%\begin{eqnarray}
=
\begin{pmatrix}
\hat{H}_{eff1}(t) & \hat{0}_{2 \times 2}\\
\hat{0}_{2 \times 2} & \hat{H}_{eff2}(t)
\end{pmatrix}, \nonumber \\
\hat{H}_{eff1}(t)=
\begin{pmatrix}
E_{c1} -E_{f1q}(t)+E_{p1} & t_s \\
t_s^{*} & E_{c1} +E_{f1q}+E_{p2} \\
\end{pmatrix}, \nonumber \\
\hat{H}_{eff2}(t)=
\begin{pmatrix}
E_{c2} -E_{f2q}(t)+E_{p1} & t_s \\
t_s^{*} & E_{c2} +E_{f2q}+E_{p2} \\
\end{pmatrix}
\end{eqnarray}
\normalsize
Here $E_{f1}(t)=a_1\frac{e(x_2-x_1)}{2}\sqrt{\frac{2}{\epsilon}\hbar \omega}  cos(\omega t)$ and $E_{f2}(t)=a_2\frac{e(x_2-x_1)}{2}\sqrt{\frac{2}{\epsilon}\hbar 2\omega} sin(2 \omega t)$ are depending on the values of 1st and 2nd energetic level. Such reasoning can be extended further for n-th energetic levels since we can write n-th oscillating cavity mode (for odd number n) as
\begin{equation}
E_{fnq}(t)=a_n\frac{e(x_2-x_1)}{2}\sqrt{\frac{2}{\epsilon}\hbar  \frac{(n+1)}{2}\omega }sin(\frac{(n+1)}{2} \omega t)
\end{equation}
and we can write n-th oscillating cavity mode (for even number n) as
\begin{equation}
E_{fnq}(t)=a_n\frac{e(x_2-x_1)}{2}\sqrt{\frac{2}{\epsilon}\hbar  \frac{(n+1)}{2}\omega }cos(\frac{(n+1)}{2} \omega t)
\end{equation}
Also one can set $E_{c1}=\frac{\hbar}{2}\omega$, $E_{c2}=\frac{3\hbar}{2}\omega$ and  $E_{cn}=\frac{\hbar(2n-1)}{2}\omega$.
Here constants $a_1, .., a_n$ dependent on the geometrical placement of position based qubit in relation to quantum electromagnetic cavity. In particular qubit can be inside or outside quantum EM cavity in the proximity of hole made in this cavity and it determines all $a_1, ..,a_n$ coefficient values.
The Hamiltonian of qubit interacting with quantum electromagnetic cavity corresponds to the following quantum state
\begin{equation}
|\psi(t)>=\gamma_1(t)|E_{c1}>|x_1>_{qubit}+\gamma_2(t)|E_{c1}>|x_2>_{qubit}+\gamma_3(t)|E_{c2}>|x_1>_{qubit}+\gamma_4(t)|E_{c2}>|x_2>_{qubit},
\end{equation}
where
$|\gamma_1|^2+|\gamma_2|^2+|\gamma_3|^2+|\gamma_4|^2=1$.
\normalsize
The equation of motion for quantum state are given by Schroedinger $\hat{H}|\psi>_t=i \hbar \frac{d}{dt}|\psi>_t$ implies the existence
\begin{eqnarray}
  |\psi(t)>=e^{(\int_{t_0}^{t}\frac{1}{\hbar i}\hat{H}(t')dt')}|\psi(t_0)>=\hat{U}(t,t_0)|\psi(t_0)>=\nonumber \\
  =e^{\frac{1}{\hbar i}\int_{t0}^{t}\begin{pmatrix}
\hat{H}_{eff1}(t') & \hat{0}_{2 \times 2}\\
\hat{0}_{2 \times 2} & \hat{H}_{eff2}(t')
  \end{pmatrix}dt'
  }|\psi(t_0)>=\nonumber \\
  =\begin{pmatrix}
  e^{\int_{t_0}^{t}\frac{1}{\hbar i}\hat{H}_{eff1}(t')dt'} & \hat{0}_{2\times2} \\
  \hat{0}_{2\times2}& e^{\int_{t_0}^{t}\frac{1}{\hbar i}\hat{H}_{eff2}(t')dt'}
  \end{pmatrix}
   |\psi(t_0)>=
   \nonumber \\ =\begin{pmatrix}
   U_{1,1}(t,t_0) & U_{1,2}(t,t_0) & 0 & 0 \\
   U_{2,1}(t,t_0) & U_{2,2}(t,t_0) & 0 & 0 \\
   0 & 0 & U_{3,3}(t,t_0) & U_{3,4}(t,t_0) \\
   0 & 0 & U_{4,3}(t,t_0) & U_{4,4}(t,t_0) \\
                  \end{pmatrix}
                  |\psi(t_0)>= \nonumber \\
                  =\begin{pmatrix}
                  U_{1,1}(t,t_0)\psi(t_0)_1+U_{1,2}(t,t_0)\psi(t_0)_2 \\
                  U_{2,1}(t,t_0)\psi(t_0)_1+U_{2,2}(t,t_0)\psi(t_0)_2 \\
                  U_{3,3}(t,t_0)\psi(t_0)_3+U_{3,4}(t,t_0)\psi(t_0)_4 \\
                  U_{4,3}(t,t_0)\psi(t_0)_3+U_{4,4}(t,t_0)\psi(t_0)_4 \\
                  \end{pmatrix}=
                  \begin{pmatrix}
                  \psi(t)_1 \\
                  \psi(t)_2 \\
                  \psi(t)_3 \\
                  \psi(t)_4 \\
                  \end{pmatrix}= |\psi(t)>.
\end{eqnarray}
In such way the density matrix of system evolution with time can be determined so it is as
\begin{eqnarray}
\hat{\rho}(t)=|\psi(t)><\psi(t)|= \nonumber \\ =
   \begin{pmatrix}
                  \psi(t)_1 \\
                  \psi(t)_2 \\
                  \psi(t)_3 \\
                  \psi(t)_4 \\
   \end{pmatrix}
     \begin{pmatrix}
                  \psi(t)_1^{*} & \psi(t)_2^{*} & \psi(t)_3^{*} & \psi(t)_4^{*}
   \end{pmatrix}= \nonumber \\ =
   \begin{pmatrix}
                  \psi(t)_1\psi(t)_1^{*} & \psi(t)_1\psi(t)_2^{*} & \psi(t)_1\psi(t)_3^{*} & \psi(t)_1\psi(t)_4^{*}  \\
                  \psi(t)_2\psi(t)_1^{*} & \psi(t)_2\psi(t)_2^{*} & \psi(t)_2\psi(t)_3^{*} & \psi(t)_2\psi(t)_4^{*}  \\
                  \psi(t)_3\psi(t)_1^{*} & \psi(t)_3\psi(t)_2^{*} & \psi(t)_3\psi(t)_3^{*} & \psi(t)_3\psi(t)_4^{*}  \\
                  \psi(t)_4\psi(t)_1^{*} & \psi(t)_4\psi(t)_2^{*} & \psi(t)_4\psi(t)_3^{*} & \psi(t)_4\psi(t)_4^{*}  \\
                 % \psi(t)_2 \\
                 % \psi(t)_3 \\
                 % \psi(t)_4 \\
   \end{pmatrix}  \nonumber \\
  % = \begin{pmatrix}
%                  \psi(t)_1\psi(t)_1^{*} & 0 & 0 & 0  \\
%                  0 & \psi(t)_2\psi(t)_2^{*} & 0 & 0  \\
%                  0 & 0 & \psi(t)_3\psi(t)_3^{*} & 0  \\
%                  0 & 0 & 0 & \psi(t)_4\psi(t)_4^{*}  \\
%                 % \psi(t)_2 \\
%                 % \psi(t)_3 \\
%                 % \psi(t)_4 \\
%   \end{pmatrix} = \nonumber \\
%   =\begin{pmatrix}
%                 0 & \psi(t)_1\psi(t)_2^{*} & \psi(t)_1\psi(t)_3^{*} & \psi(t)_1\psi(t)_4^{*}  \\
%                  \psi(t)_2\psi(t)_1^{*} & 0 & \psi(t)_2\psi(t)_3^{*} & \psi(t)_2\psi(t)_4^{*}  \\
%                  \psi(t)_3\psi(t)_1^{*} & \psi(t)_3\psi(t)_2^{*} &0 & \psi(t)_3\psi(t)_4^{*}  \\
%                  \psi(t)_4\psi(t)_1^{*} & \psi(t)_4\psi(t)_2^{*} & \psi(t)_4\psi(t)_3^{*} & 0 \\
%                 % \psi(t)_2 \\
%                 % \psi(t)_3 \\
%                 % \psi(t)_4 \\
%   \end{pmatrix} ,\nonumber \\
\end{eqnarray}
and it brings

\begin{eqnarray*}
   \begin{pmatrix}
   P(E_{c1},x_1) \\
   P(E_{c1},x_2) \\
   P(E_{c2},x_1) \\
   P(E_{c2},x_2) \\
   \end{pmatrix}=\begin{pmatrix}
                  |U(t)_{1,1}\psi(t_0)_1+U(t)_{1,2}\psi(t_0)_2|^2 \\
                  |U(t)_{2,1}\psi(t_0)_1+U(t)_{2,2}\psi(t_0)_2|^2 \\
                  |U(t)_{3,3}\psi(t_0)_3+U(t)_{3,4}\psi(t_0)_4|^2 \\
                  |U(t)_{4,3}\psi(t_0)_3+U(t)_{4,4}\psi(t_0)_4|^2 \\
                  \end{pmatrix} \times %\nonumber \\
  \end{eqnarray*}
  \tiny
   \begin{eqnarray*}
                  \times \frac{1}{|U(t)_{1,1}\psi(t_0)_1+U(t)_{1,2}\psi(t_0)_2|^2 +|U(t)_{2,1}\psi(t_0)_1+U(t)_{2,2}\psi(t_0)_2|^2+|U(t)_{3,3}\psi(t_0)_3+U(t)_{3,4}\psi(t_0)_4|^2+ |U(t)_{4,3}\psi(t_0)_3+U(t)_{4,4}\psi(t_0)_4|^2 }
\end{eqnarray*}
\normalsize
In order to determine the evolution of probabilities with time we need to establish the quantum state at initial time. By algebraic analysis of Hamiltonian we obtain 4 eigenstates and 4 eigenenergies of the system given as
\begin{eqnarray}
|v_{E1}>=\frac{1}{\sqrt{ (-2 E_{f1} + E_{p1}-E_{p2}-\sqrt{(2 E_{f1} - E_{p1} + E_{p2})^2 + 4 |t_s|^2))^2+4|t_s|^2}}} \times \nonumber \\ \times
\begin{pmatrix}
(-2 E_{f1} + E_{p1}-E_{p2}-\sqrt{(2 E_{f1} - E_{p1} + E_{p2})^2 + 4 |t_s|^2)})e^{i \alpha} \nonumber \\
2 |t_s| \nonumber \\
0 \nonumber \\
0 \nonumber \\
\end{pmatrix}
%%%%{Ecav1 + Ef1 + Ep1, E^(I alpha) ts, 0, 0}
\end{eqnarray}
\begin{eqnarray}
|v_{E2}>= \frac{1}{\sqrt{ (-2 E_{f1} + E_{p1}-E_{p2}+\sqrt{(2 E_{f1} - E_{p1} + E_{p2})^2 + 4 |t_s|^2))^2+4|t_s|^2}}}
 \times \nonumber \\ \times
\begin{pmatrix}
(-2 E_{f1} + E_{p1}-E_{p2}+\sqrt{(2 E_{f1} - E_{p1} + E_{p2})^2 + 4 |t_s|^2)})e^{i \alpha} \nonumber \\
2 |t_s| \nonumber \\
0 \nonumber \\
0 \nonumber \\
\end{pmatrix}
%%%%{Ecav1 + Ef1 + Ep1, E^(I alpha) ts, 0, 0}
\end{eqnarray}
\begin{eqnarray}
|v_{E3}>=
\frac{1}{\sqrt{ (-2 E_{f1} + E_{p1}-E_{p2}-\sqrt{(2 E_{f1} - E_{p1} + E_{p2})^2 + 4 |t_s|^2))^2+4|t_s|^2}}}
 \times \nonumber \\ \times
\begin{pmatrix}
0 \nonumber \\
0 \nonumber \\
(-2 E_{f1} + E_{p1}-E_{p2}-\sqrt{(2 E_{f1} - E_{p1} + E_{p2})^2 + 4 |t_s|^2)})e^{i \alpha} \nonumber \\
2 |t_s| \nonumber \\
\end{pmatrix}
%%%%{Ecav1 + Ef1 + Ep1, E^(I alpha) ts, 0, 0}
\end{eqnarray}
\begin{eqnarray}
%{0, 0, E^(-I alpha) ts, Ecav2 - Ef2 + Ep2
|v_{E4}>=
\frac{1}{\sqrt{ (-2 E_{f1} + E_{p1}-E_{p2}+\sqrt{(2 E_{f1} - E_{p1} + E_{p2})^2 + 4 |t_s|^2))^2+4|t_s|^2}}}
 \times \nonumber \\
\begin{pmatrix}
0 \nonumber \\
0 \nonumber \\
(-2 E_{f1} + E_{p1}-E_{p2}+\sqrt{(2 E_{f1}+ E_{p1} + E_{p2})^2 + 4 |t_s|^2)})e^{i \alpha} \nonumber \\
2 |t_s| \nonumber \\
\end{pmatrix}
%%%%{Ecav1 + Ef1 + Ep1, E^(I alpha) ts, 0, 0}
\end{eqnarray}
and as
\begin{eqnarray}
  E_1 = \frac{1}{2} (2 E_{c1} + E_{p1} + E_{p2} - \sqrt{(2 E_{f1} - E_{p1} + E_{p2})^2 + 4 |t_s|^2})\\
  E_2= \frac{1}{2} (2 E_{c1} + E_{p1} + E_{p2} + \sqrt{(2 E_{f1} - E_{p1} + E_{p2})^2 + 4 |t_s|^2}) \\
E_3=\frac{1}{2} (2 E_{c2}  + E_{p1} + E_{p2} - \sqrt{(2 E_{f2}- E_{p1} + E_{p2})^2 + 4 |t_s|^2}) \\
E_4=\frac{1}{2} (2 E_{c2}  + E_{p1} + E_{p2} + \sqrt{(2 E_{f2} - E_{p1} + E_{p2})^2 + 4 |t_s|^2})
\end{eqnarray}
All eigenenergies and eigenstates are depending on time since $E_{f1}$ and $E_{f2}$ are time-dependent.
In such way the trajectory of quantum state with time can be written as
\begin{eqnarray}
|\psi(t)>=
\frac{\sqrt{p_{E1}(t)}e^{i \phi_{E1(t)}}}{\sqrt{ (-2 E_{f1} + E_{p1}-E_{p2}-\sqrt{(2 E_{f1} - E_{p1} + E_{p2})^2 + 4 |t_s|^2))^2+4|t_s|^2}}} \times \nonumber \\ \times
\begin{pmatrix}
(-2 E_{f1} + E_{p1}-E_{p2}-\sqrt{(2 E_{f1} - E_{p1} + E_{p2})^2 + 4 |t_s|^2)})e^{i \alpha} \nonumber \\
2 |t_s| \nonumber \\
0 \nonumber \\
0 \nonumber \\
\end{pmatrix}+
 \nonumber \\
 \frac{\sqrt{p_{E2}(t)}e^{i \phi_{E2(t)}}}{\sqrt{ (-2 E_{f1} + E_{p1}-E_{p2}+\sqrt{(2 E_{f1} - E_{p1} + E_{p2})^2 + 4 |t_s|^2))^2+4|t_s|^2}}} \times \nonumber \\ \times
\begin{pmatrix}
(-2 E_{f1} + E_{p1}-E_{p2}+\sqrt{(2 E_{f1} - E_{p1} + E_{p2})^2 + 4 |t_s|^2)})e^{i \alpha} \nonumber \\
2 |t_s| \nonumber \\
0 \nonumber \\
0 \nonumber \\
\end{pmatrix}
+ \nonumber \\
+ \frac{\sqrt{p_{E3}(t)}e^{i \phi_{E3(t)}}}{\sqrt{(-2 E_{f2} + E_{p1}-E_{p2}-\sqrt{(2 E_{f1} - E_{p1} + E_{p2})^2 + 4 |t_s|^2)})^2+4|t_s|^2}} \times \nonumber \\ \times
\begin{pmatrix}
0 \nonumber \\
0 \nonumber \\
(-2 E_{f1} + E_{p1}-E_{p2}-\sqrt{(2 E_{f1} - E_{p1} + E_{p2})^2 + 4 |t_s|^2)})e^{i \alpha} \nonumber \\
2|t_s| \nonumber \\
\end{pmatrix} + \nonumber \\
+\frac{\sqrt{p_{E4}(t)}e^{i \phi_{E4(t)}}}{\sqrt{ (-2 E_{f1} + E_{p1}-E_{p2}+\sqrt{(2 E_{f1} - E_{p1} + E_{p2})^2 + 4 |t_s|^2))^2+4|t_s|^2}}} \times \nonumber \\
\begin{pmatrix}
0 \nonumber \\
0 \nonumber \\
(-2 E_{f1} + E_{p1}-E_{p2}+\sqrt{(2 E_{f1} - E_{p1} + E_{p2})^2 + 4 |t_s|^2)})e^{i \alpha} \nonumber \\
2|t_s| \nonumber \\
\end{pmatrix}
%+\gamma_2|E_{c1}>|x_2>_{qubit}+\gamma_3|E_{c2}>|x_1>_{qubit}+\gamma_4|E_{c2}>|x_2>_{qubit},
\end{eqnarray}
The measurement of state 1 at node x1 of position-based qubit is represented by the action of the operator
\begin{eqnarray}
\hat{P}_{x1}=
(|E_{c1}><E_{c1}|+|E_{c2}><E_{c2}|)|x1><x1|=
\begin{pmatrix}
1 & 0 \\
0 & 1 \\
\end{pmatrix}
\times
\begin{pmatrix}
1 & 0 \\
0 & 0 \\
\end{pmatrix}= %\nonumber \\
\begin{pmatrix}
1 & 0 & 0 & 0 \\
0 & 0 & 0 & 0 \\
0 & 0 & 1 & 0 \\
0 & 0 & 0 & 0 \\
\end{pmatrix}
\end{eqnarray} on the quantum state $|\psi>$ that is $\hat{P}_{x1}|\psi>=|\psi_1>$. %In quite analogical way we can conduct the measurement of energy occupancy in quantum EM cavity
It is instructive to observe the Rabi oscillations in the system under our consideration. First step is the
determination of density matrix, so we have
\begin{eqnarray}
\hat{\rho}_t=|\psi(t)><\psi(t)| %=\nonumber \\
=
\begin{pmatrix}
\rho_{1,1}(t) & \rho_{1,2}(t) & \rho_{1,3}(t) & \rho_{1,4}(t) \\
\rho_{1,2}^{*}(t) & \rho_{2,2}(t) & \rho_{2,3}(t) & \rho_{2,4}(t) \\
\rho_{1,3}^{*}(t) & \rho_{2,3}^{*}(t) & \rho_{3,3}(t) & \rho_{3,4}(t) \\
\rho_{1,4}^{*}(t) & \rho_{2,4}^{*}(t) & \rho_{3,4}^{*}(t) & \rho_{4,4}(t) \\
\end{pmatrix}
\end{eqnarray}
We have the probability for the electromagnetic quantum cavity to be populated by the state $|E_{c1}>$ is of the following
form
\begin{eqnarray}
P(E_{c1})_t=\rho_{1,1}(t) + \rho_{2,2}(t) %=\nonumber \\
\end{eqnarray}
and the probability for the cavity to be populated by state $|E_{c2}>$ is of the form
\begin{eqnarray}
P(E_{c2})_t=\rho_{3,3}(t) + \rho_{4,4}(t) =1-P(E_{c1})_t.
\end{eqnarray}
The probability for qubit to be in the state $|x_1>$ is of the form
\begin{eqnarray}
P(E_{x1})_t=\rho_{1,1}(t) + \rho_{3,3}(t) %=\nonumber \\
\end{eqnarray}
while the probability for qubit to be in the state $|x_2>$ is of the form
\begin{eqnarray}
P(E_{x2})_t=\rho_{2,2}(t) + \rho_{4,4}(t) %=\nonumber \\
\end{eqnarray}
The evolution with time of position based qubit can be described by density matrix that is of the form
\begin{eqnarray}
\hat{\rho(t)}_{qubit}=
\begin{pmatrix}
\rho_{1,1}(t)+\rho_{3,3}(t) & \rho_{1,2}(t)+\rho_{3,4}(t) \\
\rho_{2,1}(t)+\rho_{4,3}(t) & \rho_{2,2}(t)+\rho_{4,4}(t) \end{pmatrix}
\end{eqnarray}
In similar fashion we obtain density matrix of quantum EM cavity that is of the form
\begin{eqnarray}
\hat{\rho(t)}_{QEC}=
\begin{pmatrix}
\rho_{1,1}(t)+\rho_{2,2}(t) & \rho_{1,3}(t)+\rho_{2,4}(t) \\
\rho_{3,1}(t)+\rho_{4,2}(t) & \rho_{3,3}(t)+\rho_{4,4}(t) \end{pmatrix}=
\begin{pmatrix}
\rho_{1,1}(t)_{QEC} & \rho_{1,2}(t)_{QEC} \\
\rho_{2,1}(t)_{QEC} & \rho_{2,2}(t)_{QEC}.
\end{pmatrix}
\end{eqnarray}
The quantum entanglement between electrostatic position-based qubit and quantum cavity can be established by von Neumann entropy as given by
formula $S_{QEC}(t)=-Tr[\rho_{QEC}(t)\log(\rho_{QEC}(t))]$ that gives
\begin{eqnarray}
S_{QEC}(t)=\frac{-1}{2} (\log (1-\sqrt{(1-2 \rho_{2,2}(t)_{QEC})^2+4 |\rho_{1,2}(t)_{QEC}|^2})+\log (\sqrt{(1-2 \rho_{2,2}(t)_{QEC})^2+4 |\rho_{1,2}(t)_{QEC}|^2}+1)+ \nonumber \\
+2 \sqrt{(1-2 \rho_{2,2}(t)_{QEC})^2  %+ $ \newline $
+4|\rho_{1,2}(t)_{QEC}|^2} \tanh ^{-1}(\sqrt{(1-2\rho_{2,2}(t)_{QEC})^2  %+ \nonumber \\
+4 |\rho_{1,2}(t)_{QEC}|^2})-\log (4)).
\end{eqnarray}

%\newline
%%$+2 \sqrt{(1-2 \rho_{2,2}(t)_{QEC})^2  %+ $ \newline $
%%+4
%%   |\rho_{1,2}(t)_{QEC}|^2} \tanh ^{-1}(\sqrt{(1-2\rho_{2,2}(t)_{QEC})^2+4 |\rho_{1,2}(t)_{QEC}|^2})-\log (4))$.
Since $\rho(t)_{4 \times 4}$ is determined analytically ( as it will be shown later) it implies that structure $\rho(t)_{QEC}$ is know and this will lead to very complicated but finite length formula for entropy expressed by all Hamiltonian parameters.
\section{Case of 2 qubits interaction with electromagnetic cavity}
%Write your subsection text here.
We assume that two qubits are placed at sufficient distance one from each other so they do not interact electrostatically by Coulomb force.
If we place them in quantum electromagnetic cavity or in the proximity to the quantum electromagnetic cavity with holes they interact with
quantum activity and therefore they do indirectly interact. Existence of hole in quantum electromagnetic cavity causes the leakage of electromagnetic energy
and therefore there is escape of photons from cavity to outer space that can be accounted as quantum cavity complex value eigenfrequency. However
we omit the imaginary part as we will assume that it goes towards 0 in our simplistic model.
Total Hamiltonian accounts for energy of quantum electromagnetic cavity (QEC), energy of first and second position based qubit and for interaction between quantum electromagnetic cavity and position based qubits and is of the following form
\begin{equation}
H=H_{qubitA}+H_{qubitB}+H_{QEC}+H_{qubitA-QEC}+H_{qubitB-QEC}.
\end{equation}
It leads to the following Hamiltonian
\begin{eqnarray}
    \label{simple_equation}
    H = I_{qed} \times H_{A,qubit} \times I_{B,qubit}  + H_{qed} \times I_{A,qubit}\times I_{B,qubit} + H_{qed-Aqubit}+H_{qed-Bqubit} = \nonumber \\
%\end{eqnarray}
%\begin{eqnarray}
    =
\begin{pmatrix}
1 & 0 \\
0 & 1
\end{pmatrix} \times
\begin{pmatrix}
E_{p1} & t_s \\
t_s & E_{p2}
\end{pmatrix} \times
\begin{pmatrix}
1 & 0 \\
0 & 1
\end{pmatrix}+ \nonumber \\
+
\begin{pmatrix}
1 & 0 \\
0 & 1
\end{pmatrix} \times
\begin{pmatrix}
1 & 0 \\
0 & 1
\end{pmatrix} \times
\begin{pmatrix}
E_{p1} & t_s \\
t_s & E_{p2}
\end{pmatrix}+ \nonumber \\
+ % \nonumber \\
\begin{pmatrix}
E_{cav1} & 0 \\
0 & E_{cav2}
\end{pmatrix} \times
\begin{pmatrix}
1 & 0 \\
0 & 1
\end{pmatrix}
 \times
\begin{pmatrix}
1 & 0 \\
0 & 1
\end{pmatrix}+ \nonumber \\
+
(x_2-x_1)e
\begin{pmatrix}
E_{f1a}(t) & 0 \\
0 & E_{f2a}(t)
\end{pmatrix} \times
\begin{pmatrix}
-1 & 0 \\
0 & 1
\end{pmatrix}
\times
\begin{pmatrix}
1 & 0 \\
0 & 1
\end{pmatrix}+ \nonumber \\ +
(x_{2b}-x_{1b})e
\begin{pmatrix}
E_{f1b}(t) & 0 \\
0 & E_{f2b}(t)
\end{pmatrix}
\times
\begin{pmatrix}
1 & 0 \\
0 & 1
\end{pmatrix} \times
\begin{pmatrix}
-1 & 0 \\
0 & 1
\end{pmatrix}
%= \nonumber \\
%=
%\begin{pmatrix}
%E_{p1} & t_s & 0 & 0 \\
%t_s^{*} & E_{p2} & 0 & 0 \\
%0 & 0 & E_{p1} & t_s \\
%0 & 0 & t_s^{*} & E_{p2}
%\end{pmatrix}+ \nonumber \\ +
%\begin{pmatrix}
%E_{cav1} & 0 & 0 & 0 \\
%0 & E_{cav1} & 0 & 0 \\
%0 & 0 & E_{cav2} & 0 \\
%0 & 0 & 0 & E_{cav2}
%\end{pmatrix}+\nonumber \\
% \frac{e(x_2-x_1)}{2}
%\begin{pmatrix}
%-E_{f1} & 0  & 0 & 0 \\
%0 & E_{f1} & 0 & 0 \\
%0 & 0  & -E_{f2} & 0 \\
%0 & 0  & 0 & E_{f2}
%\end{pmatrix}= \nonumber \\
\end{eqnarray}
%\tiny
%\begin{eqnarray}
%\begin{pmatrix}
%E_{cav1} -\frac{e(x_2-x_1)}{2}E_{f1}(t)+E_{p1} & t_s  & 0 & 0 \\
%t_s^{*} & E_{cav1} +\frac{e(x_2-x_1)}{2}E_{f1}+E_{p2} & 0 & 0 \\
%0 & 0  & E_{cav2} -\frac{e(x_2-x_1)}{2}E_{f2}+E_{p1} & t_s \\
%0 & 0  & t_s^{*} & +E_{cav2}+\frac{e(x_2-x_1)}{2}E_{f2}+E_{p2}
%\end{pmatrix}
%\end{eqnarray}
Final the Hamiltonian structure for 2 qubits interacting with quantum cavity is of the following form
\begin{equation}
H=
\begin{pmatrix}
H_{eff1G}(t)_{4 \times 4} & \hat{0}_{4 \times 4} \\
\hat{0}_{4 \times 4} & H_{eff2G}(t)_{4 \times 4}
\end{pmatrix}
\end{equation}

%\right)$ \normalsize
%%%\newline
%%%\tiny
where
\begin{eqnarray*}
H_{eff1G}(t)_{4 \times 4}=
\begin{pmatrix}
 E_{c1}+E_{p1a}+E_{p1b} & e^{i \beta } \text{tsb} & e^{i \alpha } t_{sa} & 0  \\
 e^{-i \beta } t_{sb} & E_{c1}+E_{p1a}+E_{p2b} & 0 & e^{i \alpha } t_{sa}  \\
 e^{-i \alpha } t_{sa} & 0 & E_{c1}+E_{p1b}+E_{p2a} & e^{i \beta } t_{sb}  \\
 0 & e^{-i \alpha } t_{sa} & e^{-i \beta } t_{sb} & E_{c1}+E_{p2a}+E_{p2b}  \\
\end{pmatrix} + \nonumber \\
%%\end{eqnarray}
%%\begin{eqnarray}
\begin{pmatrix}
%\begin{array}{cccccccc}
 -E_{f1a}(t)-E_{f1b}(t) & 0  & 0 & 0 \\
 0 & -E_{f1a}(t)+E_{f1b}(t) & 0 & 0 \\
 0 & 0 & E_{f1a}(t)-E_{f1b}(t) & 0 \\
 0 & 0 & 0 & E_{f1a}(t)+E_{f1b}(t) \\
%\end{array}
%\right)
%$
\end{pmatrix}
\end{eqnarray*}
and
%\newline
\normalsize
\begin{eqnarray*}
H_{eff2G}(t)_{4 \times 4}=%%%%\nonumber \\
%\left(
\begin{pmatrix}
%\begin{array}{cccccccc}
 E_{c2}+E_{p1a}+E_{p1b} & e^{i \beta } t_{sb} & e^{i \alpha } t_{sa} & 0 \\
 e^{-i \beta } t_{sb} & E_{c2}+E_{p1a}+E_{p2b} & 0 & e^{i \alpha } t_{sa} \\
 e^{-i \alpha } t_{sa} & 0 & E_{c2}+E_{p1b}+E_{p2a} & e^{i \beta } t_{sb} \\
 0 & e^{-i \alpha } t_{sa} & e^{-i \beta } t_{sb} & E_{c2}+E_{p2a}+E_{p2b} \\
%\end{array}
%\right)
%$
\end{pmatrix}+ \nonumber \\
+
\begin{pmatrix}
%\begin{array}{cccccccc}
 -E_{f2a}(t)-E_{f2b}(t) & 0  & 0 & 0 \\
 0 & -E_{f2a}(t)+E_{f2b}(t) & 0 & 0 \\
 0 & 0 & E_{f2a}(t)-E_{f2b}(t) & 0 \\
 0 & 0 & 0 & E_{f2a}(t)+E_{f2b}(t) \\
%\end{array}
%\right)
%$
\end{pmatrix}
\end{eqnarray*}
%\begin{sideways}
\normalsize
It is quite straightforward to write the evolution of quantum state by
\begin{eqnarray}
  |\psi(t)>=e^{(\int_{t_0}^{t}\frac{1}{\hbar i}\hat{H}(t')dt')}|\psi(t_0)>=\hat{U}(t,t_0)|\psi(t_0)>=\nonumber \\
  =e^{\frac{1}{\hbar i}\int_{t0}^{t}\begin{pmatrix}
\hat{H}_{eff1G}(t')_{4 \times 4} & \hat{0}_{4 \times 4}\\
\hat{0}_{4 \times 4} & \hat{H}_{eff2G}(t')_{4 \times 4}
  \end{pmatrix}dt'
  }|\psi(t_0)>=\begin{pmatrix}
\hat{U1G}(t,t_0)_{4 \times 4} & \hat{0}_{4 \times 4}\\
\hat{0}_{4 \times 4} & \hat{U2G}(t,t_0)_{4 \times 4}
  \end{pmatrix}
  |\psi(t_0)>
  =\nonumber \\
  =\begin{pmatrix}
  e^{\int_{t_0}^{t}\frac{1}{\hbar i}\hat{H}_{eff1G}(t')_{4\times 4}dt'} & \hat{0}_{4\times 4} \\
  \hat{0}_{4\times 4}& e^{\int_{t_0}^{t}\frac{1}{\hbar i}\hat{H}_{eff2G}(t')_{4\times 4}dt'}
  \end{pmatrix}
   |\psi(t_0)>=
   \nonumber \\ =\begin{pmatrix}
   U_{1,1}(t,t_0) & U_{1,2}(t,t_0)& U_{1,3}(t,t_0) & U_{1,4}(t,t_0) & 0 & 0 & 0 & 0 \\
   U_{2,1}(t,t_0) & U_{2,2}(t,t_0)& U_{2,3}(t,t_0) & U_{2,4}(t,t_0) & 0 & 0 & 0 & 0 \\
   U_{3,1}(t,t_0) & U_{2,2}(t,t_0)& U_{3,3}(t,t_0) & U_{3,4}(t,t_0) & 0 & 0 & 0 & 0 \\
   U_{4,1}(t,t_0) & U_{4,2}(t,t_0)& U_{4,3}(t,t_0) & U_{4,4}(t,t_0) & 0 & 0 & 0 & 0 \\
   0 & 0 & 0 & 0 & U_{5,5}(t,t_0) & U_{5,6}(t,t_0)& U_{5,7}(t,t_0) & U_{5,8}(t,t_0) \\
   0 & 0 & 0 & 0 & U_{6,5}(t,t_0) & U_{6,6}(t,t_0)& U_{6,7}(t,t_0) & U_{6,8}(t,t_0) \\
   0 & 0 & 0 & 0 & U_{7,5}(t,t_0) & U_{7,6}(t,t_0)& U_{7,7}(t,t_0) & U_{7,8}(t,t_0) \\
   0 & 0 & 0 & 0 & U_{8,5}(t,t_0) & U_{8,6}(t,t_0)& U_{8,7}(t,t_0) & U_{8,8}(t,t_0) \\
                  \end{pmatrix}
                  |\psi(t_0)>= \nonumber \\
                  =\begin{pmatrix}
                  U_{1,1}(t,t_0)\psi(t_0)_1+U_{1,2}(t,t_0)\psi(t_0)_2+U_{1,3}(t,t_0)\psi(t_0)_3+U_{1,4}(t,t_0)\psi(t_0)_4 \\
                  U_{2,1}(t,t_0)\psi(t_0)_1+U_{2,2}(t,t_0)\psi(t_0)_2+U_{2,3}(t,t_0)\psi(t_0)_3+U_{1,4}(t,t_0)\psi(t_0)_4 \\
                  U_{3,3}(t,t_0)\psi(t_0)_3+U_{3,4}(t,t_0)\psi(t_0)_4+U_{3,3}(t,t_0)\psi(t_0)_3+U_{3,4}(t,t_0)\psi(t_0)_4 \\
                  U_{4,3}(t,t_0)\psi(t_0)_3+U_{4,4}(t,t_0)\psi(t_0)_4+U_{4,3}(t,t_0)\psi(t_0)_3+U_{4,4}(t,t_0)\psi(t_0)_4 \\
                  U_{5,5}(t,t_0)\psi(t_0)_5+U_{5,6}(t,t_0)\psi(t_0)_2+U_{5,7}(t,t_0)\psi(t_0)_7+U_{5,8}(t,t_0)\psi(t_0)_8 \\
                  U_{6,5}(t,t_0)\psi(t_0)_5+U_{6,6}(t,t_0)\psi(t_0)_6+U_{6,7}(t,t_0)\psi(t_0)_7+U_{6,8}(t,t_0)\psi(t_0)_8 \\
                  U_{7,5}(t,t_0)\psi(t_0)_5+U_{7,6}(t,t_0)\psi(t_0)_6+U_{7,7}(t,t_0)\psi(t_0)_7+U_{7,8}(t,t_0)\psi(t_0)_8 \\
                  U_{8,5}(t,t_0)\psi(t_0)_5+U_{8,6}(t,t_0)\psi(t_0)_6+U_{8,7}(t,t_0)\psi(t_0)_7+U_{8,8}(t,t_0)\psi(t_0)_8 \\
                  \end{pmatrix}=
                  \begin{pmatrix}
                  \psi(t)_1 \\
                  \psi(t)_2\\
                  \psi(t)_3 \\
                  \psi(t)_4 \\
                  \psi(t)_5 \\
                  \psi(t)_6 \\
                  \psi(t)_7 \\
                  \psi(t)_8 \\
                  \end{pmatrix}= |\psi(t)>.
\end{eqnarray}
\normalsize
%\begin{figure}
%    \centering
%    \includegraphics[width=3.0in]{myfigure}
%    \caption{Simulation Results}
%    \label{simulationfigure}
%\end{figure}
\begin{landscape}
\section{Analytical results for evolution of quantum state with time for position-based qubit interacting with quantum electromagnetic cavity}
%\begin{landscape}{angle=0}
%\tiny
\normalsize
In general case with use of 2 by 2 matrix of tight-binding model we have time-dependent parameters $t_s(t)e^{\i \alpha(t)}$ as well as $E_{p1}(t)$ and $E_{p2}(t)$. We assume that those parameters account for 2 energy level occupancy for position-based qubit. In such case the dynamics of quantum electromagnetic cavity having two occupied energy levels coupled to position based qubits can be described by $\hat{U}(t_0,t)$ matrix describing the
evolution of quantum state from initial time $t_0$ to time t. It is expressed by following 8 non-zero value coefficients given by following analytical form:
$U_{1,1}=[\exp (-\frac{1}{2\hbar}[i (\sqrt{(\int_{t_0}^{t} dt'(E_{c1}-E_{f1}(t')+E_{p1}(t')) \, dt-\int_{t_0}^{t} (E_{c1}+E_{f2}(t')+E_{p2}(t')) \, dt')^2+4 \left(\int_{t_0}^{t} e^{-i \alpha
   (t')} |t_s(t')| \, dt'\right) \int e^{i \alpha (t')} |t_s(t')| \, dt'}+$
    $+\int_{t_0}^{t} (E_{c1}-E_{f1}(t')+E_{p1(t')}) \, dt+\int_{t_0}^{t}
   (E_{c1}+E_{f2}(t')+E_{p2}(t')) \, dt')]) \times$ \newline
%ala1
    $\times ([\int_{t_0}^{t} (E_{c1}-E_{f1}(t')+E_{p1}(t')) \, dt]\times$ \newline  $\times(-(-1+\exp
   (\frac{i \sqrt{(\int (E_{c1}-E_{f1}(t')+E_{p1}(t')) \, dt-\int_{t_0}^{t} (E_{c1}+E_{f2}(t')+E_{p2}(t')) \, dt')^2+4 \left(\int_{t_0}^{t} e^{-i \alpha (t')}
   |t_s(t')| \, dt'\right) \int_{t_0}^{t} e^{i \alpha (t')} |t_s(t')| \, dt'}}{\hbar})))+$ \newline
%ala2
$+(\int (E_{c1}+E_{f2}(t')+E_{p2}(t')) \,
   dt'+$ \newline $+\sqrt{(\int_{t_0}^{t} (E_{c1}-E_{f1}(t')+E_{p1}(t')) \, dt'-\int_{t_0}^{t} (E_{c1}+E_{f2}(t')+E_{p2}(t')) \, dt')^2+4 \left(\int_{t_0}^{t} e^{-i \alpha (t')} |t_s(t')| \,
   dt'\right) \int_{t_0}^{t} e^{i \alpha (t')} |t_s(t')| \, dt'}) \times $ \newline
%ala
$ \times \exp \left(\frac{i \sqrt{(\int (E_{c1}-E_{f1}(t')+E_{p1}(t')) \, dt'-\int_{t_0}^{t}
   (E_{c1}+E_{f2}(t')+E_{p2}(t')) \, dt')^2+4 \left(\int e^{-i \alpha (t')} t_{s}(t') \, dt\right) \int e^{i \alpha (t')} |t_{s}(t')| \,
   dt'}}{\hbar}\right)+$ \newline $+\sqrt{(\int (E_{c1}-E_{f1}(t')+E_{p1}(t')) \, dt'-\int_{t_0}^{t} (E_{c1}+E_{f2}(t')+E_{p2}(t')) \, dt')^2+4 \left(\int e^{-i \alpha
   (t')} |t_s(t')| \, dt'\right) \int e^{i \alpha (t')} |t_s(t')| \, dt'}-\int_{t_0}^{t} (E_{c1}(t')+E_{f2}(t')+E_{p2}(t')) \, dt')]\times$ \newline $\times \frac{1}{2 \sqrt{(\int
   (E_{c1}-E_{f1}(t')+E_{p1}(t')) \, dt'-\int_{t_0}^{t} (E_{c1}+E_{f2}(t')+E_{p2}(t')) \, dt')^2+4 \left(\int_{t_0}^{t} e^{-i \alpha (t')} |t_s(t')| \, dt'\right) \int e^{i
   \alpha (t')} |t_s(t')| \, dt'}}$ , \newline \normalsize
   and \newline %\tiny
  $U_{2,2}(t)=[\exp (-[i (\sqrt{\left(\int_{\text{t0}}^t \left(E_{c1}-E_{f1}\left(t'\right)+E_{p1}\left(t'\right)\right) \, dt'-\int_{\text{t0}}^t
   \left(E_{c1}+E_{f2}\left(t'\right)+E_{p2}\left(t'\right)\right) \, dt'\right){}^2+4 \left(\int_{\text{t0}}^t t_{s}(t) e^{i \alpha \left(t'\right)} \,
   dt'\right) \int_{\text{t0}}^t e^{-i \alpha \left(t'\right)} t_{s}\left(t'\right) \, dt'}+$ \newline
$+\int_{\text{t0}}^t
   \left(E_{c1}-E_{f1}\left(t'\right)+E_{p1}\left(t'\right)\right) \, dt'+\int_{\text{t0}}^t
   \left(E_{c1}+E_{f2}\left(t'\right)+E_{p2}\left(t'\right)\right) \, dt')]\frac{1}{2 \hbar}) \times$ \newline
$(\sqrt{\left(\int_{\text{t0}}^t
   \left(E_{c1}-E_{f1}\left(t'\right)+E_{p1}\left(t'\right)\right) \, dt'-\int_{\text{t0}}^t
   \left(E_{c1}+E_{f2}\left(t'\right)+E_{p2}\left(t'\right)\right) \, dt'\right){}^2+4 \left(\int_{\text{t0}}^t t_{s}(t) e^{i \alpha \left(t'\right)} \,
   dt'\right) \int_{\text{t0}}^t e^{-i \alpha \left(t'\right)} t_{s}\left(t'\right) \, dt'}+$ \newline
$+ \left(1+\exp \left(\frac{i \sqrt{\left(\int_{\text{t0}}^t
   \left(E_{c1}-E_{f1}\left(t'\right)+E_{p1}\left(t'\right)\right) \, dt'-\int_{\text{t0}}^t
   \left(E_{c1}+E_{f2}\left(t'\right)+E_{p2}\left(t'\right)\right) \, dt'\right){}^2+4 \left(\int_{\text{t0}}^t t_{s}(t) e^{i \alpha \left(t'\right)} \,
   dt'\right) \int_{\text{t0}}^t e^{-i \alpha \left(t'\right)} t_{s}\left(t'\right) \, dt'}}{\text{hbar}}\right)\right)+[\int_{\text{t0}}^t
   \left(E_{c1}-E_{f1}\left(t'\right)+E_{p1}\left(t'\right)\right) \, dt']\times$ \newline $(-1+\exp \left(\frac{i \sqrt{\left(\int_{\text{t0}}^t
   \left(E_{c1}-E_{f1}\left(t'\right)+E_{p1}\left(t'\right)\right) \, dt'-\int_{\text{t0}}^t
   \left(E_{c1}+E_{f2}\left(t'\right)+E_{p2}\left(t'\right)\right) \, dt'\right){}^2+4 \left(\int_{\text{t0}}^t t_{s}(t) e^{i \alpha \left(t'\right)} \,
   dt'\right) \int_{\text{t0}}^t e^{-i \alpha \left(t'\right)} t_{s}\left(t'\right) \, dt'}}{\hbar}\right))$ \newline
$-[\int_{\text{t0}}^t
   \left(E_{c1}+E_{f2}\left(t'\right)+E_{p2}\left(t'\right)\right) \, dt']\times $ \newline
 $(-1+\exp \left(\frac{i \sqrt{\left(\int_{\text{t0}}^t
   \left(E_{c1}-E_{f1}\left(t'\right)+E_{p1}\left(t'\right)\right) \, dt'-\int_{\text{t0}}^t
   \left(E_{c1}+E_{f2}\left(t'\right)+E_{p2}\left(t'\right)\right) \, dt'\right){}^2+4 \left(\int_{\text{t0}}^t t_{s}(t) e^{i \alpha \left(t'\right)} \,
   dt'\right) \int_{\text{t0}}^t e^{-i \alpha \left(t'\right)} t_{s}\left(t'\right) \, dt'}}{\hbar}\right)))]\times $ \newline $ \times
\frac{1}{2 \sqrt{\left(\int_{\text{t0}}^t
   \left(E_{c1}-E_{f1}\left(t'\right)+E_{p1}\left(t'\right)\right) \, dt'-\int_{\text{t0}}^t
   \left(E_{c1}+E_{f2}\left(t'\right)+E_{p2}\left(t'\right)\right) \, dt'\right){}^2+4 \left(\int_{\text{t0}}^t t_{s}(t) e^{i \alpha \left(t'\right)} \,
   dt'\right) \int_{\text{t0}}^t e^{-i \alpha \left(t'\right)} t_{s}\left(t'\right) \, dt'}}$ \normalsize
 and %\tiny
 \newline
 $U_{3,3}(t)=[\exp (-[i (\sqrt{\left(\int_{\text{t0}}^t \left(E_{c2}-E_{f1}\left(t'\right)+E_{p1}\left(t'\right)\right) \, dt'-\int_{\text{t0}}^t
   \left(E_{c2}+E_{f2}\left(t'\right)+E_{p2}\left(t'\right)\right) \, dt'\right){}^2+4 \left(\int_{\text{t0}}^t e^{-i \alpha \left(t'\right)}
   t_{s}\left(t'\right) \, dt'\right) \int_{\text{t0}}^t e^{i \alpha \left(t'\right)} t_{s}\left(t'\right) \, dt'}+$ \newline
$+\int_{\text{t0}}^t
   \left(E_{c2}-E_{f1}\left(t'\right)+E_{p1}\left(t'\right)\right) \, dt'+\int_{\text{t0}}^t
   \left(E_{c2}+E_{f2}\left(t'\right)+E_{p2}\left(t'\right)\right) \, dt')] \frac{1}{2 \hbar})$ \newline  $(\sqrt{\left(\int_{\text{t0}}^t
   \left(E_{c2}-E_{f1}\left(t'\right)+E_{p1}\left(t'\right)\right) \, dt'-\int_{\text{t0}}^t
   \left(E_{c2}+E_{f2}\left(t'\right)+E_{p2}\left(t'\right)\right) \, dt'\right){}^2+4 \left(\int_{\text{t0}}^t e^{-i \alpha \left(t'\right)}
   t_{s}\left(t'\right) \, dt'\right) \int_{\text{t0}}^t e^{i \alpha \left(t'\right)} t_{s}\left(t'\right) \, dt'} $
\newline
$\left(1+\exp \left(\frac{i
   \sqrt{\left(\int_{\text{t0}}^t \left(E_{c2}-E_{f1}\left(t'\right)+E_{p1}\left(t'\right)\right) \, dt'-\int_{\text{t0}}^t
   \left(E_{c2}+E_{f2}\left(t'\right)+E_{p2}\left(t'\right)\right) \, dt'\right){}^2+4 \left(\int_{\text{t0}}^t e^{-i \alpha \left(t'\right)}
   t_{s}\left(t'\right) \, dt'\right) \int_{\text{t0}}^t e^{i \alpha \left(t'\right)} t_{s}\left(t'\right) \, dt'}}{\hbar}\right)\right)-[\int_{\text{t0}}^t
   \left(E_{c2}-E_{f1}\left(t'\right)+E_{p1}\left(t'\right)\right) \, dt' ]\times $ \newline
$(-1+\exp (i \sqrt{\left(\int_{\text{t0}}^t
   \left(E_{c2}-E_{f1}\left(t'\right)+E_{p1}\left(t'\right)\right) \, dt'-\int_{\text{t0}}^t
   \left(E_{c2}+E_{f2}\left(t'\right)+E_{p2}\left(t'\right)\right) \, dt'\right){}^2+4 \left(\int_{\text{t0}}^t e^{-i \alpha \left(t'\right)}
   t_{s}\left(t'\right) \, dt'\right) \int_{\text{t0}}^t e^{i \alpha \left(t'\right)} t_{s}\left(t'\right) \, dt'})\frac{1}{\hbar}))+\int_{\text{t0}}^t
   \left(E_{c2}+E_{f2}\left(t'\right)+E_{p2}\left(t'\right)\right) \, dt' $ \newline
$\left(-1+\exp \left(\frac{i \sqrt{\left(\int_{\text{t0}}^t
   \left(E_{c2}-E_{f1}\left(t'\right)+E_{p1}\left(t'\right)\right) \, dt'-\int_{\text{t0}}^t
   \left(E_{c2}+E_{f2}\left(t'\right)+E_{p2}\left(t'\right)\right) \, dt'\right){}^2+4 \left(\int_{\text{t0}}^t e^{-i \alpha \left(t'\right)}
   t_{s}\left(t'\right) \, dt'\right) \int_{\text{t0}}^t e^{i \alpha \left(t'\right)} t_{s}\left(t'\right) \, dt'}}{\hbar}\right)\right))]\times $ \newline $\times \frac{1}{2
   \sqrt{\left(\int_{\text{t0}}^t \left(E_{c2}-E_{f1}\left(t'\right)+E_{p1}\left(t'\right)\right) \, dt'-\int_{\text{t0}}^t
   \left(E_{c2}+E_{f2}\left(t'\right)+E_{p2}\left(t'\right)\right) \, dt'\right){}^2+4 \left(\int_{\text{t0}}^t e^{-i \alpha \left(t'\right)}
   t_{s}\left(t'\right) \, dt'\right) \int_{\text{t0}}^t e^{i \alpha \left(t'\right)} t_{s}\left(t'\right) \, dt'}}$
   \newline
\normalsize
   and
%\tiny
   \newline
   $U_{4,4}(t)=
   [\exp (-[i(\sqrt{\left(\int_{t_{0}}^t \left(E_{c2}-E_{f1}\left(t'\right)+E_{p1}\left(t'\right)\right) \, dt'-\int_{\text{t0}}^t
   \left(E_{c2}+E_{f2}\left(t'\right)+E_{p2}\left(t'\right)\right) \, dt'\right){}^2+4 \left(\int_{\text{t0}}^t e^{-i \alpha \left(t'\right)}
   t_{s}\left(t'\right) \, dt'\right) \int_{\text{t0}}^t e^{i \alpha \left(t'\right)} t_{s}\left(t'\right) \, dt'}+\int_{\text{t0}}^t
   \left(E_{c2}-E_{f1}\left(t'\right)+E_{p1}\left(t'\right)\right) \, dt'+\int_{\text{t0}}^t
   \left(E_{c2}+E_{f2}\left(t'\right)+E_{p2}\left(t'\right)\right) \, dt')]\frac{1}{2 \hbar}) $ \newline
$(\sqrt{\left(\int_{\text{t0}}^t
   \left(E_{c2}-E_{f1}\left(t'\right)+E_{p1}\left(t'\right)\right) \, dt'-\int_{\text{t0}}^t
   \left(E_{c2}+E_{f2}\left(t'\right)+E_{p2}\left(t'\right)\right) \, dt'\right){}^2+4 \left(\int_{\text{t0}}^t e^{-i \alpha \left(t'\right)}
   t_{s}\left(t'\right) \, dt'\right) \int_{\text{t0}}^t e^{i \alpha \left(t'\right)} t_{s}\left(t'\right) \, dt'}$ \newline
 $\left(1+\exp \left(\frac{i
   \sqrt{\left(\int_{\text{t0}}^t \left(E_{c2}-E_{f1}\left(t'\right)+E_{p1}\left(t'\right)\right) \, dt'-\int_{\text{t0}}^t
   \left(E_{c2}+E_{f2}\left(t'\right)+E_{p2}\left(t'\right)\right) \, dt'\right){}^2+4 \left(\int_{\text{t0}}^t e^{-i \alpha \left(t'\right)}
   t_{s}\left(t'\right) \, dt'\right) \int_{\text{t0}}^t e^{i \alpha \left(t'\right)} t_{s}\left(t'\right) \, dt'}}{\hbar}\right)\right)+$
\newline
 $+[\int_{\text{t0}}^t
   \left(E_{c2}-E_{f1}\left(t'\right)+E_{p1}\left(t'\right)\right) \, dt']$ \newline
$ \left(-1+\exp \left(\frac{i \sqrt{\left(\int_{\text{t0}}^t
   \left(E_{c2}-E_{f1}\left(t'\right)+E_{p1}\left(t'\right)\right) \, dt'-\int_{\text{t0}}^t
   \left(E_{c2}+E_{f2}\left(t'\right)+E_{p2}\left(t'\right)\right) \, dt'\right){}^2+4 \left(\int_{\text{t0}}^t e^{-i \alpha \left(t'\right)}
   t_{s}\left(t'\right) \, dt'\right) \int_{\text{t0}}^t e^{i \alpha \left(t'\right)} t_{s}\left(t'\right) \, dt'}}{\hbar}\right)\right)$ \nonumber
$-[\int_{\text{t0}}^t
   \left(E_{c2}+E_{f2}\left(t'\right)+E_{p2}\left(t'\right)\right) \, dt']$ \newline
 $\left(-1+\exp \left(\frac{i \sqrt{\left(\int_{\text{t0}}^t
   \left(E_{c2}-E_{f1}\left(t'\right)+E_{p1}\left(t'\right)\right) \, dt'-\int_{\text{t0}}^t
   \left(E_{c2}+E_{f2}\left(t'\right)+E_{p2}\left(t'\right)\right) \, dt'\right){}^2+4 \left(\int_{\text{t0}}^t e^{-i \alpha \left(t'\right)}
   t_{s}\left(t'\right) \, dt'\right) \int_{\text{t0}}^t e^{i \alpha \left(t'\right)} t_{s}\left(t'\right) \, dt'}}{\hbar}\right)\right))] \times$

$ \frac{1}{2
   \sqrt{\left(\int_{\text{t0}}^t \left(E_{c2}-E_{f1}\left(t'\right)+E_{p1}\left(t'\right)\right) \, dt'-\int_{\text{t0}}^t
   \left(E_{c2}+E_{f2}\left(t'\right)+E_{p2}\left(t'\right)\right) \, dt'\right){}^2+4 \left(\int_{\text{t0}}^t e^{-i \alpha \left(t'\right)}
   t_{s}\left(t'\right) \, dt'\right) \int_{\text{t0}}^t e^{i \alpha \left(t'\right)} t_{s}\left(t'\right) \, dt'}}
   $ \newline \normalsize and \newline %\tiny
   $U_{1,2}(t)=-[(\int_{\text{t0}}^t t_{s}(t) e^{i \alpha \left(t'\right)} \, dt')\times$ \newline $\times \exp (-[i (\sqrt{\left(\int_{\text{t0}}^t
   \left(E_{c1}-E_{f1}\left(t'\right)+E_{p1}\left(t'\right)\right) \, dt'-\int_{\text{t0}}^t
   \left(E_{c1}+E_{f2}\left(t'\right)+E_{p2}\left(t'\right)\right) \, dt'\right){}^2+4 \left(\int_{\text{t0}}^t t_{s}(t) e^{i \alpha \left(t'\right)} \,
   dt'\right) \int_{\text{t0}}^t e^{-i \alpha \left(t'\right)} t_{s}\left(t'\right) \, dt'}+$ \newline $+\int_{\text{t0}}^t
   \left(E_{c1}-E_{f1}\left(t'\right)+E_{p1}\left(t'\right)\right) \, dt'+\int_{\text{t0}}^t
   \left(E_{c1}+E_{f2}\left(t'\right)+E_{p2}\left(t'\right)\right) \, dt')]\frac{1}{2 \hbar})\times$ \newline $\times (-1+\exp \left(\frac{i \sqrt{\left(\int_{\text{t0}}^t
   \left(E_{c1}-E_{f1}\left(t'\right)+E_{p1}\left(t'\right)\right) \, dt'-\int_{\text{t0}}^t
   \left(E_{c1}+E_{f2}\left(t'\right)+E_{p2}\left(t'\right)\right) \, dt'\right){}^2+4 \left(\int_{\text{t0}}^t t_{s}(t) e^{i \alpha \left(t'\right)} \,
   dt'\right) \int_{\text{t0}}^t e^{-i \alpha \left(t'\right)} t_{s}\left(t'\right) \, dt'}}{\hbar}\right))]\times$ \newline $\times\frac{1}{\sqrt{\left(\int_{\text{t0}}^t
   \left(E_{c1}-E_{f1}\left(t'\right)+E_{p1}\left(t'\right)\right) \, dt'-\int_{\text{t0}}^t
   \left(E_{c1}+E_{f2}\left(t'\right)+E_{p2}\left(t'\right)\right) \, dt'\right){}^2+4 \left(\int_{\text{t0}}^t t_{s}(t) e^{i \alpha \left(t'\right)} \,
   dt'\right) \int_{\text{t0}}^t e^{-i \alpha \left(t'\right)} t_{s}\left(t'\right) \, dt'}}$ \newline \normalsize
   and
%\tiny
%%   and
 \newline

   $U_{2,1}(t)=-[(\int_{\text{t0}}^t e^{-i \alpha \left(t'\right)} t_{s}\left(t'\right) \, dt')\times$ \newline $\times \exp (-[i (\sqrt{\left(\int_{\text{t0}}^t
   \left(E_{c1}-E_{f1}\left(t'\right)+E_{p1}\left(t'\right)\right) \, dt'-\int_{\text{t0}}^t
   \left(E_{c1}+E_{f2}\left(t'\right)+E_{p2}\left(t'\right)\right) \, dt'\right){}^2+4 \left(\int_{\text{t0}}^t t_{s}(t) e^{i \alpha \left(t'\right)} \,
   dt'\right) \int_{\text{t0}}^t e^{-i \alpha \left(t'\right)} t_{s}\left(t'\right) \, dt'}+ $ \newline $+\int_{\text{t0}}^t
   \left(E_{c1}-E_{f1}\left(t'\right)+E_{p1}\left(t'\right)\right) \, dt'+\int_{\text{t0}}^t
   \left(E_{c1}+E_{f2}\left(t'\right)+E_{p2}\left(t'\right)\right) \, dt')]\frac{1}{2 \hbar})\times$ \newline $\times \left(-1+\exp \left(\frac{i \sqrt{\left(\int_{\text{t0}}^t
   \left(E_{c1}-E_{f1}\left(t'\right)+E_{p1}\left(t'\right)\right) \, dt'-\int_{\text{t0}}^t
   \left(E_{c1}+E_{f2}\left(t'\right)+E_{p2}\left(t'\right)\right) \, dt'\right){}^2+4 \left(\int_{\text{t0}}^t t_{s}(t) e^{i \alpha \left(t'\right)} \,
   dt'\right) \int_{\text{t0}}^t e^{-i \alpha \left(t'\right)} t_{s}\left(t'\right) \, dt'}}{\hbar}\right)\right)]\times$ \newline $\times\frac{1}{\sqrt{\left(\int_{\text{t0}}^t
   \left(E_{c1}-E_{f1}\left(t'\right)+E_{p1}\left(t'\right)\right) \, dt'-\int_{\text{t0}}^t
   \left(E_{c1}+E_{f2}\left(t'\right)+E_{p2}\left(t'\right)\right) \, dt'\right){}^2+4 \left(\int_{\text{t0}}^t t_{s}(t) e^{i \alpha \left(t'\right)} \,
   dt'\right) \int_{\text{t0}}^t e^{-i \alpha \left(t'\right)} t_{s}\left(t'\right) \, dt'}}$ \newline and \newline
   $U_{3,4}(t)=-[(\int_{\text{t0}}^t e^{i \alpha \left(t'\right)} t_{s}\left(t'\right) \, dt')\times$ \newline $\times \exp (-(i (\sqrt{\left(\int_{\text{t0}}^t
   \left(E_{c2}-E_{f1}\left(t'\right)+E_{p1}\left(t'\right)\right) \, dt'-\int_{\text{t0}}^t
   \left(E_{c2}+E_{f2}\left(t'\right)+E_{p2}\left(t'\right)\right) \, dt'\right){}^2+4 \left(\int_{\text{t0}}^t e^{-i \alpha \left(t'\right)}
   t_{s}\left(t'\right) \, dt'\right) \int_{\text{t0}}^t e^{i \alpha \left(t'\right)} t_{s}\left(t'\right) \, dt'}+$ \newline $+\int_{\text{t0}}^t
   \left(E_{c2}-E_{f1}\left(t'\right)+E_{p1}\left(t'\right)\right) \, dt'+\int_{\text{t0}}^t
   \left(E_{c2}+E_{f2}\left(t'\right)+E_{p2}\left(t'\right)\right) \, dt'))\frac{1}{2 \hbar})\times$ \newline $\times \left(-1+\exp \left(\frac{i \sqrt{\left(\int_{\text{t0}}^t
   \left(E_{c2}-E_{f1}\left(t'\right)+E_{p1}\left(t'\right)\right) \, dt'-\int_{\text{t0}}^t
   \left(E_{c2}+E_{f2}\left(t'\right)+E_{p2}\left(t'\right)\right) \, dt'\right){}^2+4 \left(\int_{\text{t0}}^t e^{-i \alpha \left(t'\right)}
   t_{s}\left(t'\right) \, dt'\right) \int_{\text{t0}}^t e^{i \alpha \left(t'\right)} t_{s}\left(t'\right) \,
   dt'}}{\hbar}\right)\right)]\times $ \newline $\times \frac{1}{\sqrt{\left(\int_{\text{t0}}^t \left(E_{c2}-E_{f1}\left(t'\right)+E_{p1}\left(t'\right)\right) \, dt'-\int_{\text{t0}}^t
   \left(E_{c2}+E_{f2}\left(t'\right)+E_{p2}\left(t'\right)\right) \, dt'\right){}^2+4 \left(\int_{\text{t0}}^t e^{-i \alpha \left(t'\right)}
   t_{s}\left(t'\right) \, dt'\right) \int_{\text{t0}}^t e^{i \alpha \left(t'\right)} t_{s}\left(t'\right) \, dt'}}$
 \newline %and
\normalsize
   and
%\tiny
 \newline
 $U_{4,3}(t)=-[[\int_{\text{t0}}^t e^{-i \alpha \left(t'\right)} t_{s}\left(t'\right) \, dt'] \times $ \newline $ \exp (-[i (\sqrt{\left(\int_{\text{t0}}^t
   \left(E_{c2}\left(t'\right)-E_{f1}\left(t'\right)+E_{p1}\left(t'\right)\right) \, dt'-\int_{\text{t0}}^t
   \left(E_{c2}\left(t'\right)+E_{f2}\left(t'\right)+E_{p2}\left(t'\right)\right) \, dt'\right){}^2+4 \left(\int_{\text{t0}}^t e^{-i \alpha \left(t'\right)}
   t_{s}\left(t'\right) \, dt'\right) \int_{\text{t0}}^t e^{i \alpha \left(t'\right)} t_{s}\left(t'\right) \, dt'}+\int_{\text{t0}}^t
   \left(E_{c2}\left(t'\right)-E_{f1}\left(t'\right)+E_{p1}\left(t'\right)\right) \, dt'+\int_{\text{t0}}^t
   \left(E_{c2}\left(t'\right)+E_{f2}\left(t'\right)+E_{p2}\left(t'\right)\right) \, dt')]\frac{1}{2 \hbar})\times$ \newline $\times \left(-1+\exp \left(\frac{i
   \sqrt{\left(\int_{\text{t0}}^t \left(E_{c2}\left(t'\right)-E_{f1}\left(t'\right)+E_{p1}\left(t'\right)\right) \, dt'-\int_{\text{t0}}^t
   \left(E_{c2}\left(t'\right)+E_{f2}\left(t'\right)+E_{p2}\left(t'\right)\right) \, dt'\right){}^2+4 \left(\int_{\text{t0}}^t e^{-i \alpha \left(t'\right)}
   t_{s}\left(t'\right) \, dt'\right) \int_{\text{t0}}^t e^{i \alpha \left(t'\right)} t_{s}\left(t'\right) \,
   dt'}}{\hbar}\right)\right)]\times$ \newline $\times\frac{1}{\sqrt{\left(\int_{\text{t0}}^t \left(E_{c2}\left(t'\right)-E_{f1}\left(t'\right)+E_{p1}\left(t'\right)\right) \,
   dt'-\int_{\text{t0}}^t \left(E_{c2}\left(t'\right)+E_{f2}\left(t'\right)+E_{p2}\left(t'\right)\right) \, dt'\right){}^2+4 \left(\int_{\text{t0}}^t e^{-i \alpha
   \left(t'\right)} t_{s}\left(t'\right) \, dt'\right) \int_{\text{t0}}^t e^{i \alpha \left(t'\right)} t_{s}\left(t'\right) \, dt'}}$
   \normalsize
  \end{landscape}
%\end{landscape}
\normalsize
\section{Towards N energetic levels of quantum electromagnetic cavity interacting with K energetic levels of position-based qubit}
It is natural to generalize our considerations for the case of quantum electromagnetic cavity with N energetic levels and qubit with K energetic levels in the framework of Jaynes-Cummings tight-binding Hamiltonian. Let us set for simplicity N=4 and K=4. We have the following Hamiltonian

\begin{eqnarray*}
  %  \label{simple_equation}
    H = I_{QEC} \times H_{qubit,E1-E2} \times I_{qubit,E3-E4}  +\nonumber \\ + H_{QEC} \times I_{qubit,E1-E2}\times I_{qubit,E3-E4} + H_{QEC-qubit-E1-E2}+H_{QEC-qubit-E3-E4} = \nonumber \\
%\end{eqnarray}
%\begin{eqnarray}
    =
\begin{pmatrix}
1 & 0 & 0 & 0 \\
0 & 1 & 0 & 0 \\
0 & 0 & 1 & 0 \\
0 & 0 & 0 & 1
\end{pmatrix} \times
\begin{pmatrix}
E_{p1}(t)_{Eq1,Eq2} & t_{s,Eq1,Eq2} \\
t_{s,Eq1,Eq2}^{*} & E_{p2}(t)_{s,Eq1,Eq2}
\end{pmatrix} \times
\begin{pmatrix}
1 & 0 \\
0 & 1
\end{pmatrix}+ \nonumber \\
+
\begin{pmatrix}
1 & 0 & 0 & 0\\
0 & 1 & 0 & 0\\
0 & 0 & 1 & 0\\
0 & 0 & 0 & 1\\
\end{pmatrix} \times
\begin{pmatrix}
1 & 0 \\
0 & 1
\end{pmatrix} \times
\begin{pmatrix}
E_{p1}(t)_{Eq3,Eq4} & t_{s,Eq3,Eq4} \\
t_{s,Eq3,Eq4}^{*} & E_{p2}(t)_{Eq3,Eq4}
\end{pmatrix}+ \nonumber \\
+ % \nonumber \\
\begin{pmatrix}
E_{c1} & 0 & 0 & 0 \\
0 & E_{c2}& 0 & 0 \\
0 & 0 & E_{c3} & 0 \\
0 & 0 & 0 & E_{c4} \\
\end{pmatrix} \times
\begin{pmatrix}
1 & 0 \\
0 & 1
\end{pmatrix}
 \times
\begin{pmatrix}
1 & 0 \\
0 & 1
\end{pmatrix}+ \nonumber \\
+
(x_2-x_1)e
\begin{pmatrix}
E_{f1-Eq1-Eq2}(t) & 0 & 0 & 0 \\
0 & E_{f2-Eq1-Eq2}(t) & 0 & 0 \\
0 & 0 & E_{f3-Eq1-Eq2}(t) & 0 \\
0 & 0 & 0 & E_{f4-Eq1-Eq2}(t) \\
\end{pmatrix} \times
\begin{pmatrix}
-1 & 0 \\
0 & +1
\end{pmatrix}
\times
\begin{pmatrix}
1 & 0 \\
0 & 1
\end{pmatrix}+ \nonumber \\ +
(x_{2b}-x_{1b})e
\begin{pmatrix}
E_{f1-Eq3-Eq4}(t) & 0 & 0 & 0 \\
0 & E_{f2-Eq3-Eq4}(t) & 0 & 0 \\
0 & 0 & E_{f3-Eq3-Eq4}(t) & 0 \\
0 & 0 & 0 & E_{f4-Eq3-Eq4}(t) \\
\end{pmatrix}
\times
\begin{pmatrix}
1 & 0 \\
0 & 1
\end{pmatrix} \times
\begin{pmatrix}
-1 & 0 \\
0 & +1
\end{pmatrix}
%= \nonumber \\
%=
%\begin{pmatrix}
%E_{p1} & t_s & 0 & 0 \\
%t_s^{*} & E_{p2} & 0 & 0 \\
%0 & 0 & E_{p1} & t_s \\
%0 & 0 & t_s^{*} & E_{p2}
%\end{pmatrix}+ \nonumber \\ +
%\begin{pmatrix}
%E_{cav1} & 0 & 0 & 0 \\
%0 & E_{cav1} & 0 & 0 \\
%0 & 0 & E_{cav2} & 0 \\
%0 & 0 & 0 & E_{cav2}
%\end{pmatrix}+\nonumber \\
% \frac{e(x_2-x_1)}{2}
%\begin{pmatrix}
%-E_{f1} & 0  & 0 & 0 \\
%0 & E_{f1} & 0 & 0 \\
%0 & 0  & -E_{f2} & 0 \\
%0 & 0  & 0 & E_{f2}
%\end{pmatrix}= \nonumber \\
\end{eqnarray*}
Our Hamiltonian corresponds to the quantum state given as
\begin{eqnarray}
|\psi>_t=\gamma_1(t)|E_{c1}>(|x_1>_{E1-E2}|x_1>_{E3-E4})_{qubit}+\gamma_2(t)|E_{c1}>(|x_1>_{E1-E2}|x_2>_{E3-E4})_{qubit}+ \nonumber \\
+\gamma_3(t)|E_{c1}>(|x_2>_{E1-E2}|x_1>_{E3-E4})_{qubit}+\gamma_4(t)|E_{c1}>(|x_2>_{E1-E2}|x_2>_{E3-E4})_{qubit}+ \nonumber \\
+\gamma_5(t)|E_{c2}>(|x_1>_{E1-E2}|x_1>_{E3-E4})_{qubit}+\gamma_6(t)|E_{c2}>(|x_1>_{E1-E2}|x_2>_{E3-E4})_{qubit}+ \nonumber \\
+\gamma_7(t)|E_{c2}>(|x_2>_{E1-E2}|x_1>_{E3-E4})_{qubit}+\gamma_8(t)|E_{c2}>(|x_2>_{E1-E2}|x_2>_{E3-E4})_{qubit}+ \nonumber \\
+\gamma_9(t)|E_{c3}>(|x_1>_{E1-E2}|x_1>_{E3-E4})_{qubit}+\gamma_{10}(t)|E_{c3}>(|x_1>_{E1-E2}|x_2>_{E3-E4})_{qubit}+ \nonumber \\
+\gamma_{11}(t)|E_{c3}>(|x_2>_{E1-E2}|x_1>_{E3-E4})_{qubit}+\gamma_{12}(t)|E_{c3}>(|x_2>_{E1-E2}|x_2>_{E3-E4})_{qubit}+ \nonumber \\
+\gamma_{13}(t)|E_{c4}>(|x_1>_{E1-E2}|x_1>_{E3-E4})_{qubit}+\gamma_{14}(t)|E_{c4}>(|x_1>_{E1-E2}|x_2>_{E3-E4})_{qubit}+ \nonumber \\
+\gamma_{15}(t)|E_{c4}>(|x_2>_{E1-E2}|x_1>_{E3-E4})_{qubit}+\gamma_{16}(t)|E_{c4}>(|x_2>_{E1-E2}|x_2>_{E3-E4})_{qubit}, \nonumber \\
\end{eqnarray}
with normalization condition $|\gamma_1|^2+..+|\gamma_{16}|^2=1$.
In the next step we compute evolution operator that is the sum of tensor products of Pauli matrices. Such evolution operator has the analytical form and it gives us the analytic form the quantum trajectory of the system. At this stage the numerics is only needed for illustration of obtained analytical solutions.
Once $\hat{U}$ is determined we can establish mulitphoton processes. Let us assume that in initial state the cavity has populated $E_{c1}$ state.
It means that
\begin{eqnarray}
|\psi>_{t0}=\gamma_1(t_0)|E_{c1}>(|x_1>_{E1-E2}|x_1>_{E3-E4})_{qubit}+\gamma_2(t_0)|E_{c1}>(|x_1>_{E1-E2}|x_2>_{E3-E4})_{qubit}+ \nonumber \\
+\gamma_3(t_0)|E_{c1}>(|x_2>_{E1-E2}|x_1>_{E3-E4})_{qubit}+\gamma_4(t_0)|E_{c1}>(|x_2>_{E1-E2}|x_2>_{E3-E4})_{qubit}. % \nonumber \\
\end{eqnarray}
After certain time t the system is in new state $|\psi>_t$. If we want to establish the probability for effective 3 photon processes we need to act with operator
\begin{eqnarray}
P_{Ec4}=(|E_{c4}><E_{c4}|)\times \hat{I}_{E1-E2} \times \hat{I}_{E3-E4}=
\begin{pmatrix}
0 & 0 & 0 & 0 \\
0 & 0 & 0 & 0 \\
0 & 0 & 0 & 0 \\
0 & 0 & 0 & 1 \\
\end{pmatrix}\times \begin{pmatrix}
1 & 0 & 0 & 0 \\
0 & 1 & 0 & 0 \\
0 & 0 & 1 & 0 \\
0 & 0 & 0 & 1 \\
\end{pmatrix}
.
\end{eqnarray}
Probability for occurrence of multi-photon processes so probability of transition from quantum cavity eigenstate $E_{c1}$ to $E_{c4}$ is given by formula
\begin{eqnarray}
Prob(E_{c1} \rightarrow E_{c4})=|<\psi|_{t0}P_{Ec4}|\psi>_{t}|^2.
\end{eqnarray}

\section{Conclusion}
The whole dynamics of position-based qubit interacting with quantum cavity was established in analytical way. The whole dynamics of qubit and quantum EM cavity density matrices dependence with time is obtained. It allows in characterization of Rabi oscillations in analytical way.
At the same time we can obtain analytical formula for dependence of quantum von-Neumann entropy with time for position-based qubit interacting with electromagnetic cavity coupled to other qubits. It is quite straightforward to generalize the obtained results for N qubits interacting with electromagnetic quantum cavity.
Quantum state evolution operator is always expressed by the analytic and elementary functions and it eliminates the need in usage of numerical approach in determination of quantum state evolution. The described work can be generalized further for various configurations of interacting qubit clusters interacting with quantum electromagnetic cavity. 
% Since in such case the time evolution operator is deterministic and analytical we have deterministic evolution of N qubits interacting with electromagnetic cavity.
\section{Acknowledgment}
 The assistance in picture preparation was done by Erik Staszewski from University College Dublin.

\end{document}